# Relative Effect of Inclinations for Moonlets in the Triple Asteroidal Systems

Yu Jiang[1, 2], Hexi Baoyin[2] , Yun Zhang[2]

1. State Key Laboratory of Astronautic Dynamics, Xi'an Satellite Control Center, Xi'an 710043, China

2. School of Aerospace Engineering, Tsinghua University, Beijing 100084, China

Y. Jiang (🖂) e-mail: jiangyu_xian_china@163.com (corresponding author)

**Abstract**. We present the analysis and computational results for the inclination relative effect of moonlets of triple asteroidal systems. Perturbations on moonlets due to the primary's non-sphericity gravity, the solar gravity, and moonlets' relative gravity are discussed. The inclination vector for each moonlet follows a periodic elliptical motion; the motion period depends on the moonlet's semi-major axis and the primary's J2 perturbations. Perturbation on moonlets from the Solar gravity and moonlet's relative gravity makes the motion of the x component of the inclination vector of moonlet 1 and the y component of the inclination vector of moonlet 2 to be periodic.The mean motion of x component and the y component of the inclination vector of each moonlet forms an ellipse. However, the instantaneous motion of x component and the y component of the inclination vector may be an elliptical disc due to the coupling effect of perturbation forces. Furthermore, the x component of the inclination vector of moonlet 1 and the y component of the inclination vector of moonlet 2 form a quasi-periodic motion. Numerical calculation of dynamical configurations of two triple asteroidal systems (216) Kleopatra and (153591) 2001 SN263 validates the conclusion.

**Key words**: triple asteroidal system; minor celestial bodies; nonlinear dynamics;

## 1. Introduction

To study the dynamical mechanism of triple asteroidal systems can not only help us to understand the origin of the Solar system and the formation of the asteroidal belt (Araujo et al. 2012), but also help to design the orbit of spacecraft in the human's future space mission to triple asteroidal systems. The first triple asteroid (87) Sylvia was discovered in 2005 (Marchis et al. 2005), after that, there are eight such triple asteroidal systems and one Kuiper-belt object discovered in the solar system. Table 1 shows the physical and orbital parameters of these triple asteroidal systems. Two of them are trinary near-Earth-Asteroid systems (NEAs), i.e. 136617 1994CC (Brozović et al. 2011; Fang et al. 2011) and 153591 2001SN263 (Fang et al. 2011; Araujo et al. 2012). Besides, 47171 1999TC36 (Benecchi et al. 2010) and 136108 Haumea (Pinilla-Alonso et al. 2009; Lockwood et al. 2014) are trans-Neptunian objects (TNOs). Others are main-belt triple asteroidal systems.

Table 1. Physical and orbital parameters of triple asteroidal systems

| Name of triple asteroid system | Primary | | | Diameters of primary, second component, and third component (km) |
|---|---|---|---|---|
| | Mass (kg) | Bulk density ($g \cdot cm^{-3}$) | Rotation period (h) | |
| (45) Eugenia[a1-a3] | $5.62887\times10^{18}$ | 1.1 | 5.699 | 304×220×146, 5, 7 |
| (87) Sylvia[b1-b5] | $1.478\times10^{19}$ | 1.29 | 5.18364 | 385×262×232, 10.8, 10.6 |
| (93) Minerva[c1-c3] | $3.35\times10^{18}$ | 1.75 | 5.981767 | 141.6, 3.6, 3.2 |
| (216) Kleopatra[d1-d8] | $4.64\times10^{18}$ | 3.6 | 5.385 | 217×94×81, 8.9, 6.9 |
| 3749 Balam[e] | $5.1\times10^{14}$ | 2.6 | 2.80483 | 3.95, 1.84, 1.66 |
| 47171 1999TC36[f1,f2] | $1.28\times10^{19}$ | 0.64 | 45.763 | 272, 132, 251 |
| 136108 Haumea[g1-g3] | $4.03\times10^{21}$ | 2.97 | 3.9154 | 1379, 320, 160 |
| (136617) 1994CC[h1,h2] | $2.66\times10^{11}$ | 2.1 | 2.3886 | 0.69×0.67×0.64, 0.113, 0.08 |
| 153591 2001SN263[i1-i3] | $9.51\times10^{12}$ | 1.1±0.2 | 3.4256±0.0002 | 2.5±0.3, 0.77±0.12, 0.43±0.14 |

[a1]Beauvalet et al. 2012. [a2]Beauvalet and Marchis 2014. [a3]Marchis et al. 2010. [b1]Berthier et al. 2014. [b2]Fang et al. 2012. [b3]Frouard et al. 2012. [b4]Marchis et al. 2005. [b5]Winter et al. 2009. [c1]Marchis et al. 2011. [c2]Marchis et al. 2013. [c3]Torppa et al. 2008. [d1]Descamps et al. 2011. [d2]Jiang and Baoyin 2014. [d3]Jiang et al. 2014. [d4]Jiang 2015. [d5]Jiang et al. 2015a. [d6]Jiang et al. 2015b. [d7]Jiang et al. 2015c. [d8]Ostro et al. 2000. [e]Vokrouhlický 2009. [f1]Benecchi et al. 2010. [f2]Mommert et al. 2012. [g1]Dumas et al. 2011. [g2]Pinilla-Alonso et al. 2009. [g3]Lockwood et al. 2014. [h1]Brozović et al. 2011. [h2]Fang et al. 2011. [i1]Fang et al. 2011. [i2]Araujo et al. 2012. [i3]Becker et al. 2015

The calculation of dynamical parameters of triple asteroidal systems is the basis for the study of dynamical mechanism for these systems. Marchis et al. (2005) presented the two moonlets of (87) Sylvia orbiting at 710 and 1,360 km, and the J2 of the two moonlet are 0.17 and 0.18, respectively. Ragozzine and Brown (2009) studied the orbits and masses of satellites of 136108 Haumea and indicated that Haumea could have experienced a great collision billions of years ago. Marchis et al. (2010) found that the inclinations of moonlets of (45) Eugenia are quite different from other known main-belt triple asteroidal systems, the inclinations of the two moonlets Petit-Prince and Princesse relative to the primary's equator, are $9°$ and $18°$, respectively. Fang et al. (2012) found that the moonlets of (87) Sylvia orbiting at $807.5 \pm 2.5$km and $1357 \pm 4.0$km, and the inclinations are $7.824°$ and $8.293°$, respectively. Marchis et al. (2013) investigated the triple asteroidal system (93) Minerva and found that the moonlets of (93) Minerva are 3km and 4km in diameter, respectively. Beauvalet and Marchis (2014) analyzed the J2 of two triple asteroidal systems (45) Eugenia and (87) Sylvia, and derived the internal structure of these two triple systems. Jiang et al. (2015a) found that the number and position of equilibrium points around the primary of (216) Kleopatra will vary while the rotational speed of the primary change.

The study of dynamical behaviours of triple asteroidal systems includes orbital elements, spin-orbit lock, bifurcations, resonance, stable and unstable regions, etc. Winter et al. (2009) indicated that the longitude of the orbital nodes of the two moonlets of (87) Sylvia, Romulus and Remus, are locked to each other. Brozović et al. (2011) found that the inner moonlet of (136617) 1994CC is spin-orbit locked relative to the primary and the outer moonlet is not spin-orbit locked. Fang et al. (2011) calculated the motion of moonlets of (153591) 2001SN263 and (136617) 1994CC, examined the mean-motion resonance, Kozai resonance, and evection resonance for these two triple asteroidal systems, the results illustrated that the moonlets are not in these three resonance cases. Araujo et al. (2012) investigated the stable region of the three components of (153591) 2001SN263, they divided the region around (153591) 2001SN263 into four distinct regions and found that the stable regions are near Alpha and Beta while resonance motion with Beta and Gamma are unstable. Fang et al. (2012) deduced that the (87) Sylvia is not in the 8:3 mean-motion resonance, besides, they calculated the effects of a pass through 3:1 mean-motion eccentricity-type resonance. Frouard and Compère (2012) studied the instability zones for moonlets of the triple asteroidal system (87) Sylvia with considering the non-sphericity of Sylvia, and found that this triple system is in a deeply stable zone. Marchis et al. (2013) found that the moonlets of Minerva are at 1% and 2% of the Hill radius. Jiang et al. (2015b) found four kinds of bifurcations of periodic orbit families in the potential of the primary of (216) Kleopatra. Araujo et al. (2015) considered a massless particle in the vicinity of (153591) 2001SN263 and found that the stable regions of the particle's retrograde orbits are much bigger that the prograde orbits.

Using the perturbation method, the motion of the moonlets relative to the primary of the large size ratio triple asteroid system can be analyzed. Kozai (1959) derived the perturbations of orbital elements of a satellite in the gravitational potential of the Earth. Cook (1962) presented the perturbations from the Sun and Moon to the orbital elements of a satellite in the gravitational potential of the Earth. Allan (1970) discussed the critical inclination with the $J_2$ and $J_4$ term. For the orbits with small inclinations, the orbital element can be indicated with the inclination vectors (Hiztz 2008). The perturbation method can be applied to analyze the motion of moonlets relative to the primary in the binary and triple asteroid systems. Araujo et al. (2015) found that the $J_2$ term of the primary has a significant effect to the stable retrograde orbits in the triple asteroid 2001 SN263.

In this work we focus on the moonlets' relative effect in the triple asteroidal systems. In Section 2, the perturbation on the two moonlets due to the Solar's gravity and the primary's non-sphericity gravity are derived, and then the relative perturbation effects between these two moonlets have been investigated. In Section 3, the primary's $J_2$, Solar gravity, and the two moonlets' relative effect are all considered to analyze the dynamical system of the inclination vectors of the two moonlets. We find that for each moonlet, the inclination vector forms a periodic elliptical motion.

## 2. Perturbation on Moonlets Due to the Solar Gravity and the Primary's Non-sphericity Gravity

In this section, we derive the formulas of perturbation on moonlets due to the solar gravity and the primary's non-sphericity gravity. Denote $J_2$ as the value of the primary's $J_2$ perturbation, $G$ as the Newtonian gravitational constant, $m_{major}$ as the mass of the primary, $\mu = Gm_{major}$, $r$ as the primary's mean radius. Let $a$ be the semi-major axis, $n = \sqrt{\frac{\mu}{a^3}}$ as the mean orbit angular speed, $e$ be the eccentricity, $i$ be the inclination, $\Omega$ be the longitude of the ascending node, $\omega$ be the argument of periapsis, $M$ be the mean anomaly, $m$ be the mass. Denote the inclination vector $\begin{cases} i_x = i\sin\Omega \\ i_y = i\cos\Omega \end{cases}$. The subscripts M1, M2, and s represent orbital parameters of Moonlet 1, Moonlet 2 and Sun, respectively. Denote $\sigma_{M1} = \dfrac{m_{M1}}{m_{M1} + m_{major}}$ and $\sigma_{M2} = \dfrac{m_{M2}}{m_{M2} + m_{major}}$.

## 2.1 Perturbation on Moonlets Due to the Primary's Non-sphericity Gravity

Consider the primary's $J_2$ perturbation acting on the two moonlets, the rates of average change (Kozai 1959) of inclination and right ascension of the ascending node are

$$\begin{cases} \dfrac{di}{dt} = 0 \\ \dfrac{d\Omega}{dt} = \dfrac{3nJ_2\mu r^2}{2a^2\left(1-e^2\right)^2}\cos i \end{cases}. \tag{1}$$

For the orbits with small inclination, use the Lagrange's planetary equations (Cook 1962), we have

$$\begin{cases} \dfrac{di_x}{dt} = \sin\Omega\dfrac{di}{dt} + i\cos\Omega\dfrac{d\Omega}{dt} \\ \dfrac{di_y}{dt} = \cos\Omega\dfrac{di}{dt} - i\sin\Omega\dfrac{d\Omega}{dt} \end{cases}. \tag{2}$$

Substituting Eq. (1) into Eq. (2) and using small angle approximations, then the inclination vector's secular variation for moonlet 1 can be expressed as

$$\begin{cases} \dfrac{di_{x-M1}}{dt} = -\dfrac{3J_2\mu r^2}{2n_{M1}a_{M1}^5}i_{y-M1} \\ \dfrac{di_{y-M1}}{dt} = \dfrac{3J_2\mu r^2}{2n_{M1}a_{M1}^5}i_{x-M1} \end{cases}, \tag{3}$$

and the inclination vector's secular variation for moonlet 2 can be expressed as

$$\begin{cases} \dfrac{di_{x-M2}}{dt} = -\dfrac{3J_2\mu r^2}{2n_{M2}a_{M2}^5} i_{y-M2} \\ \dfrac{di_{y-M2}}{dt} = \dfrac{3J_2\mu r^2}{2n_{M2}a_{M2}^5} i_{x-M2} \end{cases} . \tag{4}$$

where $n_{M1}$ and $n_{M2}$ are mean orbit angular speed for moonlet 1 and moonlet 2, respectively. $i_{x-M1}$ and $i_{y-M1}$ are components of inclination vector of moonlet 1, while $i_{x-M2}$ and $i_{y-M2}$ are components of inclination vector of moonlet 2. $a_{M1}$ and $a_{M2}$ are semi-major axes for moonlet 1 and moonlet 2, respectively.

These two equations can be rewritten by

$$\frac{d}{dt}\begin{bmatrix} i_{x-M1} \\ i_{y-M1} \end{bmatrix} = K_1 \begin{bmatrix} i_{x-M1} \\ i_{y-M1} \end{bmatrix}, \tag{5}$$

and

$$\frac{d}{dt}\begin{bmatrix} i_{x-M2} \\ i_{y-M2} \end{bmatrix} = K_2 \begin{bmatrix} i_{x-M2} \\ i_{y-M2} \end{bmatrix}, \tag{6}$$

where

$$K_1 = \begin{pmatrix} 0 & -\dfrac{3J_2\mu r^2}{2n_{M1}a_{M1}^5} \\ \dfrac{3J_2\mu r^2}{2n_{M1}a_{M1}^5} & 0 \end{pmatrix} \text{ and } K_2 = \begin{pmatrix} 0 & -\dfrac{3J_2\mu r^2}{2n_{M2}a_{M2}^5} \\ \dfrac{3J_2\mu r^2}{2n_{M2}a_{M2}^5} & 0 \end{pmatrix}. \tag{7}$$

Eigenvalues of $K_1$ are $\pm\dfrac{3J_2\mu r^2}{2n_{M1}a_{M1}^5} j$ while eigenvalues of $K_2$ are $\pm\dfrac{3J_2\mu r^2}{2n_{M2}a_{M2}^5} j$ , where $j = \sqrt{-1}$. Thus we can conclude that the primary's $J_2$ perturbation make each moonlet's inclination vector to be periodic motion. The motion trajectory of the extremal point of the inclination vector is an ellipse. The motion periods are $\dfrac{4\pi n_{M1}a_{M1}^5}{3J_2\mu r^2}$ and $\dfrac{4\pi n_{M2}a_{M2}^5}{3J_2\mu r^2}$, respectively.

## 2.2 Perturbation on Moonlets Due to the Solar Gravity and Moonlet's Relative Gravity

Here we only consider the solar gravity and moonlet's relative gravity. The rates of average change of inclination and right ascension of the ascending node due to the third body's gravity (Cook 1962) are

$$\begin{cases} \dfrac{di}{dt} = \dfrac{3}{2}\dfrac{\eta}{n}\alpha\gamma \\ \dfrac{d\Omega}{dt} = \dfrac{3}{2}\dfrac{\eta}{n\sin i}\beta\gamma \end{cases}, \tag{8}$$

where

$$\eta = \frac{GM_d}{r_d^3} \tag{9}$$

and

$$\begin{cases} \alpha = \cos(\Omega-\Omega_d)\cos u_d + \cos i_d \sin u_d \sin(\Omega-\Omega_d) \\ \beta = \left[-\sin(\Omega-\Omega_d)\cos u_d + \cos i_d \sin u_d \cos(\Omega-\Omega_d)\right]\cos i + \sin i \sin i_d \sin u_d \\ \gamma = \left[\sin(\Omega-\Omega_d)\cos u_d - \cos i_d \sin u_d \cos(\Omega-\Omega_d)\right]\sin i + \cos i \sin i_d \sin u_d \end{cases}. \tag{10}$$

Here the subscript d represents orbital parameters of the third body. $u = \omega + f$, $f$ is the true anomaly.

The solar gravity and moonlet's relative gravity acting on the inclination vector of moonlet 1 is (the derivation is presented in appendix A)

$$\begin{cases} \dfrac{di_{x-M1}}{dt} = \dfrac{3}{4}n_{M1}\left(\dfrac{n_s}{n_{M1}}\right)^2 \sin^2\beta_s \sin 2i_s + \dfrac{3}{8}\sigma_{M2}n_{M1}\left(\dfrac{n_{M2}}{n_{M1}}\right)^2 \sin 2i_{M2}\cos\Omega_{M2} \\ \dfrac{di_{y-M1}}{dt} = \dfrac{3}{4}n_{M1}\left(\dfrac{n_s}{n_{M1}}\right)^2 \sin 2\beta_s \sin i_s - \dfrac{3}{8}\sigma_{M2}n_{M1}\left(\dfrac{n_{M2}}{n_{M1}}\right)^2 \sin 2i_{M2}\sin\Omega_{M2} \end{cases}, \tag{11}$$

where $n_s$ represents the mean orbit angular speed for the Sun in the primary's centroid inertial coordinate system, which equals to the triple asteroidal system's mean orbit angular speed relative to the Sun; $\beta_s$ and $i_s$ represent the true anomaly and the inclination of the Sun in the primary's centroid inertial coordinate system, respectively. $i_{M2}$ and $\Omega_{M2}$ represent the inclination and the longitude of the ascending node of moonlet 2 in the primary's centroid inertial coordinate system, respectively.

Consider the secular item, one can easily obtain

$$\begin{cases} \dfrac{di_{x-M1}}{dt} = \dfrac{3}{8} n_{M1} \left( \dfrac{n_s}{n_{M1}} \right)^2 \sin 2i_s + \dfrac{3}{8} \sigma_{M2} n_{M1} \left( \dfrac{n_{M2}}{n_{M1}} \right)^2 \sin 2i_{M2} \cos \Omega_{M2} \\ \dfrac{di_{y-M1}}{dt} = -\dfrac{3}{8} \sigma_{M2} n_{M1} \left( \dfrac{n_{M2}}{n_{M1}} \right)^2 \sin 2i_{M2} \sin \Omega_{M2} \end{cases}, \qquad (12)$$

Where $\overline{\sin^2 \beta_s} = \dfrac{1}{2}$ and $\overline{\sin 2\beta_s} = 0$ is applied to the above equation.

In like manner, the Solar gravity and moonlet's relative gravity acting on the inclination vector of moonlet 2 is

$$\begin{cases} \dfrac{di_{x-M2}}{dt} = \dfrac{3}{8} n_{M2} \left( \dfrac{n_s}{n_{M2}} \right)^2 \sin 2i_s + \dfrac{3}{8} \sigma_{M1} n_{M2} \left( \dfrac{n_{M1}}{n_{M2}} \right)^2 \sin 2i_{M1} \cos \Omega_{M1} \\ \dfrac{di_{y-M2}}{dt} = -\dfrac{3}{8} \sigma_{M1} n_{M2} \left( \dfrac{n_{M1}}{n_{M2}} \right)^2 \sin 2i_{M1} \sin \Omega_{M1} \end{cases}, \qquad (13)$$

where $i_{M1}$ and $\Omega_{M1}$ represent the inclination and the longitude of the ascending node of moonlet 1 in the primary's centroid inertial coordinate system, respectively.

Consider the moonlets are in the orbit which is near the equator of the primary. This assumption is satisfied for most of the triple asteroidal systems (Beauvalet and Marchis 2014; Fang et al. 2012; Descamps et al. 2011; Vokrouhlický 2009). With this assumption, in Eq. (12), one have

$$\begin{cases} \sin 2i_{M2}\cos\Omega_{M2} = 2\sin i_{M2}\cos i_{M2}\cos\Omega_{M2} = 2i_{y-M2}\cos i_{M2} = 2i_{y-M2}\left(1+O\left(i_{M2}^2\right)\right) \simeq 2i_{y-M2} \\ \sin 2i_{M2}\sin\Omega_{M2} = 2\sin i_{M2}\cos i_{M2}\sin\Omega_{M2} = 2i_{x-M2}\cos i_{M2} = 2i_{x-M2}\left(1+O\left(i_{M2}^2\right)\right) \simeq 2i_{x-M2} \end{cases} \tag{14}$$

Substituting Eq. (14) into Eq. (12) yields the following equation

$$\begin{cases} \dfrac{di_{x-M1}}{dt} = \dfrac{3}{8} n_{M1}\left(\dfrac{n_s}{n_{M1}}\right)^2 \sin 2i_s + \dfrac{3}{4}\sigma_{M2} n_{M1}\left(\dfrac{n_{M2}}{n_{M1}}\right)^2 i_{y-M2} \\ \dfrac{di_{y-M1}}{dt} = -\dfrac{3}{4}\sigma_{M2} n_{M1}\left(\dfrac{n_{M2}}{n_{M1}}\right)^2 i_{x-M2} \end{cases}. \tag{15}$$

In the same way, we have

$$\begin{cases} \dfrac{di_{x-M2}}{dt} = \dfrac{3}{8} n_{M2}\left(\dfrac{n_s}{n_{M2}}\right)^2 \sin 2i_s + \dfrac{3}{4}\sigma_{M1} n_{M2}\left(\dfrac{n_{M1}}{n_{M2}}\right)^2 i_{y-M1} \\ \dfrac{di_{y-M2}}{dt} = -\dfrac{3}{4}\sigma_{M1} n_{M2}\left(\dfrac{n_{M1}}{n_{M2}}\right)^2 i_{x-M1} \end{cases}. \tag{16}$$

From Eq. (15) and Eq. (16), we obtain two planar dynamical systems

$$\begin{cases} \dfrac{di_{x-M1}}{dt} = \dfrac{3}{8} n_{M1}\left(\dfrac{n_s}{n_{M1}}\right)^2 \sin 2i_s + \dfrac{3}{4}\sigma_{M2} n_{M1}\left(\dfrac{n_{M2}}{n_{M1}}\right)^2 i_{y-M2} \\ \dfrac{di_{y-M2}}{dt} = -\dfrac{3}{4}\sigma_{M1} n_{M2}\left(\dfrac{n_{M1}}{n_{M2}}\right)^2 i_{x-M1} \end{cases}, \tag{17}$$

and

$$\begin{cases} \dfrac{di_{x-M2}}{dt} = \dfrac{3}{8} n_{M2}\left(\dfrac{n_s}{n_{M2}}\right)^2 \sin 2i_s + \dfrac{3}{4}\sigma_{M1} n_{M2}\left(\dfrac{n_{M1}}{n_{M2}}\right)^2 i_{y-M1} \\ \dfrac{di_{y-M1}}{dt} = -\dfrac{3}{4}\sigma_{M2} n_{M1}\left(\dfrac{n_{M2}}{n_{M1}}\right)^2 i_{x-M2} \end{cases}. \tag{18}$$

Eq. (17) indicates that the x component of the inclination vector of moonlet 1 and the y component of the inclination vector of moonlet 2 form a planar dynamical system, while Eq. (18) indicates that the x component of the inclination vector of moonlet 2 and the y component of the inclination vector of moonlet 1 form a planar dynamical system. These two planar dynamical systems can be expressed as Eq. (19) and Eq. (21)

$$\frac{d}{dt}\begin{bmatrix} i_{x-M1} \\ i_{y-M2} \end{bmatrix} = A\begin{bmatrix} i_{x-M1} \\ i_{y-M2} \end{bmatrix} + B, \tag{19}$$

where

$$A = \begin{pmatrix} 0 & \frac{3}{4}\sigma_{M2} n_{M1} \left(\frac{n_{M2}}{n_{M1}}\right)^2 \\ -\frac{3}{4}\sigma_{M1} n_{M2} \left(\frac{n_{M1}}{n_{M2}}\right)^2 & 0 \end{pmatrix} \text{ and } B = \begin{bmatrix} \frac{3}{8} n_{M1} \left(\frac{n_s}{n_{M1}}\right)^2 \sin 2i_s \\ 0 \end{bmatrix}. \tag{20}$$

$$\frac{d}{dt}\begin{bmatrix} i_{x-M2} \\ i_{y-M1} \end{bmatrix} = C\begin{bmatrix} i_{x-M2} \\ i_{y-M1} \end{bmatrix} + D, \tag{21}$$

where

$$C = \begin{pmatrix} 0 & \frac{3}{4}\sigma_{M1} n_{M2} \left(\frac{n_{M1}}{n_{M2}}\right)^2 \\ -\frac{3}{4}\sigma_{M2} n_{M1} \left(\frac{n_{M2}}{n_{M1}}\right)^2 & 0 \end{pmatrix} \text{ and } D = \begin{bmatrix} \frac{3}{8} n_{M2} \left(\frac{n_s}{n_{M2}}\right)^2 \sin 2i_s \\ 0 \end{bmatrix}. \tag{22}$$

Using the theory from Strogatz (1994, see page 150-151), for a two-dimensional nonlinear system, the linear stability of the system can be determined by the linearized system. The linearized system of Eq. (19) is $\frac{d}{dt}\begin{bmatrix} i_{x-M1} \\ i_{y-M2} \end{bmatrix} = A\begin{bmatrix} i_{x-M1} \\ i_{y-M2} \end{bmatrix}$, The Jacobian matrix is $A$, which is a constant matrix. Eigenvalues of $A$ are $\pm\frac{3}{4}\left(\sigma_{M1}\sigma_{M2} n_{M1} n_{M2}\right)^{\frac{1}{2}} j$, which means that the planar dynamical system Eq. (19) is linearly stable. Eigenvalues of $C$ are also $\pm\frac{3}{4}\left(\sigma_{M1}\sigma_{M2} n_{M1} n_{M2}\right)^{\frac{1}{2}} j$, which means that the planar dynamical system Eq. (21) is also linearly stable.

Let $K=\frac{3}{4}\sqrt{\sigma_{M1}\sigma_{M2}n_{M1}n_{M2}}$ , then Eq. (19) and Eq. (21) can also be expressed as

$$\begin{cases}\frac{d^2 i_{x-M1}}{dt^2}=-K^2 i_{x-M1}\\ \frac{d^2 i_{y-M2}}{dt^2}=-\frac{9}{32}\sigma_{M1}n_{M2}n_{M1}\left(\frac{n_s}{n_{M2}}\right)^2\sin 2i_s-K^2 i_{y-M2}\end{cases}, \tag{23}$$

and

$$\begin{cases}\frac{d^2 i_{x-M2}}{dt^2}=-K^2 i_{x-M2}\\ \frac{d^2 i_{y-M1}}{dt^2}=-\frac{9}{32}\sigma_{M2}n_{M1}n_{M2}\left(\frac{n_s}{n_{M1}}\right)^2\sin 2i_s-K^2 i_{y-M1}\end{cases}. \tag{24}$$

The form of Eq. (23) and Eq. (24) looks like the equation of harmonic oscillator which has no frictional damping. For instance, $-K^2 i_{y-M2}$ is like the linear restoring force in the harmonic oscillator. The frequency is $K=\frac{3}{4}\sqrt{\sigma_{M1}\sigma_{M2}n_{M1}n_{M2}}$ , and the period is $T=\frac{2\pi}{K}$. The motion of the x component of the inclination vector of moonlet 1 and the y component of the inclination vector of moonlet 2 is periodic. The motion x component of the inclination vector of moonlet 2 and the y component of the inclination vector of moonlet 1 is periodic.

## 3. Relative Effect on Inclination Vectors Between the two Moonlets

In this section, the primary's $J_2$, Solar gravity, and the two moonlets' relative effect are all calculated. Consider the primary's $J_2$, Solar gravity, and moonlet 2 gravity acting on the inclination vector of moonlet 1 as well as moonlet 1 gravity acting on the inclination vector of moonlet 2, then combine Eqs. (3), (4), (17), and (18), one can obtain the following equation

$$\begin{cases} \dfrac{di_{x-M1}}{dt} = -\dfrac{3J_2\mu r^2}{2n_{M1}a_{M1}^5} i_{y-M1} + \dfrac{3}{8} n_{M1} \left( \dfrac{n_s}{n_{M1}} \right)^2 \sin 2i_s + \dfrac{3}{4} \sigma_{M2} n_{M1} \left( \dfrac{n_{M2}}{n_{M1}} \right)^2 i_{y-M2} \\ \dfrac{di_{y-M1}}{dt} = \dfrac{3J_2\mu r^2}{2n_{M1}a_{M1}^5} i_{x-M1} - \dfrac{3}{4} \sigma_{M2} n_{M1} \left( \dfrac{n_{M2}}{n_{M1}} \right)^2 i_{x-M2} \\ \dfrac{di_{x-M2}}{dt} = -\dfrac{3J_2\mu r^2}{2n_{M2}a_{M2}^5} i_{y-M2} + \dfrac{3}{8} n_{M2} \left( \dfrac{n_s}{n_{M2}} \right)^2 \sin 2i_s + \dfrac{3}{4} \sigma_{M1} n_{M2} \left( \dfrac{n_{M1}}{n_{M2}} \right)^2 i_{y-M1} \\ \dfrac{di_{y-M2}}{dt} = \dfrac{3J_2\mu r^2}{2n_{M2}a_{M2}^5} i_{x-M2} - \dfrac{3}{4} \sigma_{M1} n_{M2} \left( \dfrac{n_{M1}}{n_{M2}} \right)^2 i_{x-M1} \end{cases} . \tag{25}$$

This equation can be simplified into

$$\frac{d}{dt} \begin{bmatrix} i_{x-M1} \\ i_{y-M1} \\ i_{x-M2} \\ i_{y-M2} \end{bmatrix} = E \begin{bmatrix} i_{x-M1} \\ i_{y-M1} \\ i_{x-M2} \\ i_{y-M2} \end{bmatrix} + F , \tag{26}$$

where

$$E = \begin{bmatrix} 0 & -\dfrac{3J_2\mu r^2}{2n_{M1}a_{M1}^5} & 0 & \dfrac{3}{4} \sigma_{M2} n_{M1} \left( \dfrac{n_{M2}}{n_{M1}} \right)^2 \\ \dfrac{3J_2\mu r^2}{2n_{M1}a_{M1}^5} & 0 & -\dfrac{3}{4} \sigma_{M2} n_{M1} \left( \dfrac{n_{M2}}{n_{M1}} \right)^2 & 0 \\ 0 & \dfrac{3}{4} \sigma_{M1} n_{M2} \left( \dfrac{n_{M1}}{n_{M2}} \right)^2 & 0 & -\dfrac{3J_2\mu r^2}{2n_{M2}a_{M2}^5} \\ -\dfrac{3}{4} \sigma_{M1} n_{M2} \left( \dfrac{n_{M1}}{n_{M2}} \right)^2 & 0 & \dfrac{3J_2\mu r^2}{2n_{M2}a_{M2}^5} & 0 \end{bmatrix} \tag{27}$$

$$F = \begin{bmatrix} \dfrac{3}{8} n_{M1} \left( \dfrac{n_s}{n_{M1}} \right)^2 \sin 2i_s \\ 0 \\ \dfrac{3}{8} n_{M2} \left( \dfrac{n_s}{n_{M2}} \right)^2 \sin 2i_s \\ 0 \end{bmatrix} . \tag{28}$$

For the triple asteroidal systems, the influence on the inclination vector from the primary's $J_2$ perturbation is bigger than from the Solar gravity and the two moonlets' relative effect. The Solar gravity and the two moonlets' relative effect make the x component of the inclination vector of moonlet 1 and the y component of the inclination vector of moonlet 2 form a planar dynamical system, meanwhile, they make the x component of the inclination vector of moonlet 2 and the y component of the inclination vector of moonlet 1 form a planar dynamical system. However, the primary's $J_2$ perturbation make the x component and the y component of the inclination vector of moonlet 1 form a planar dynamical system, meanwhile, it makes the x component and the y component of the inclination vector of moonlet 2 form a planar dynamical system.

Generally speaking, for the inclination vector, the influence from the $J_2$ perturbation of the primary is much bigger than from the Solar gravity and the two moonlets' relative effect. This implies that the mean motion of x component and the y component of the inclination vector of each moonlet forms an ellipse; however, the instantaneous motion of x component and the y component of the inclination vector of each moonlet may form an elliptical disc. In addition, the x component of the inclination vector of moonlet 1 and the y component of the inclination vector of moonlet 2 form a quasi-periodic motion, and the x component of the inclination vector of moonlet 2 and the y component of the inclination vector of moonlet 1 form a quasi-periodic motion.

Two triple asteroidal systems, (216) Kleopatra and (153591) 2001 SN263 are taken as examples to verify the above theory. The gravitational field and irregular shape of the primary is computed with shape model data using the polyhedral model (Neese 2004). The primary's gravitational potential (Werner 1994; Werner and Scheeres 1997) can be computed by

$$U = \frac{1}{2} G\sigma \sum_{e \in edges} \mathbf{r}_e \bullet \mathbf{E}_e \bullet \mathbf{r}_e \cdot L_e - \frac{1}{2} G\sigma \sum_{f \in faces} \mathbf{r}_f \bullet \mathbf{F}_f \bullet \mathbf{r}_f \cdot \omega_f \ , \tag{29}$$

the primary's gravitational force is calculated by

$$\nabla U = -G\sigma \sum_{e \in edges} \mathbf{E}_e \bullet \mathbf{r}_e \cdot L_e + G\sigma \sum_{f \in faces} \mathbf{F}_f \bullet \mathbf{r}_f \cdot \omega_f \ , \tag{30}$$

while the Hessian matrix of the primary's gravitational potential can be calculated by

$$\nabla(\nabla U) = G\sigma \sum_{e \in edges} \mathbf{E}_e \cdot L_e - G\sigma \sum_{f \in faces} \mathbf{F}_f \cdot \omega_f \ , \tag{31}$$

where $G=6.67\times10^{-11} m^3 kg^{-1} s^{-2}$ represents the Newtonian gravitational constant, $\sigma$ represents the primary's bulk density; $\mathbf{r}_e$ and $\mathbf{r}_f$ are body-fixed vectors, $\mathbf{r}_e$ is from the field point to the point on the edge $e$ while $\mathbf{r}_f$ is from the field point to the point on the face $f$; $\mathbf{E}_e$ and $\mathbf{F}_f$ are geometric parameters, $\mathbf{E}_e$ is related to edges while $\mathbf{F}_f$ is related to faces; $L_e$ is the integration factor while $\omega_f$ is the solid angle.

We apply the above results to two triple asteroidal system (216) Kleopatra and (153591) 2001 SN263. Moonlets of (216) Kleopatra are Alexhelios and Cleoselene, while moonlets of (153591) 2001 SN263 are Beta and Gamma. Table 2 shows the initial orbital parameters for the moonlets of two triple asteroidal systems used in the calculation. To compare with the theoretical results of the previous contents, we use the gravitational model and integrate the dynamical equation to calculate the inclination vectors. The dynamical equations are

$$\begin{cases} \dot{\mathbf{p}}_k = \mathbf{f}_k \\ \dot{\mathbf{r}}_k = \dfrac{\mathbf{p}_k}{m_k} \\ \dot{\mathbf{K}}_k = \mathbf{n}_k \\ \dot{\mathbf{A}}_k = \hat{\boldsymbol{\psi}}_k \mathbf{A}_k \end{cases}, \qquad k = 1,2,3, \tag{32}$$

where $\boldsymbol{\psi}_k = \mathbf{A}_k \mathbf{I}_k^{-1} \mathbf{A}_k^{\mathrm{T}} \left( \mathbf{K}_k - \mathbf{r}_k \times \mathbf{p}_k \right)$, $\mathbf{r}_k$ represents the position vector of the $k$-th body, $\mathbf{p}_k = m_k \dot{\mathbf{r}}_k$ represents the linear momentum vector, $\mathbf{f}_k$ represents the gravitational force acting on the $k$-th body, $\mathbf{K}_k$ represents the angular momentum vector, $\mathbf{A}_k$ is the attitude matrix. $\mathbf{n}_k$ is the resultant gravitational torque acting on the $k$-th body. All the vectors are expressed in the inertial space. $\hat{\boldsymbol{\psi}}_k$ is calculated with the following method. For a vector $\mathbf{v} = \left[ v_x, v_y, v_z \right]^{\mathrm{T}}$, define the matrix

$$\hat{\mathbf{v}} = \begin{pmatrix} 0 & -v_z & v_y \\ v_z & 0 & -v_x \\ -v_y & v_x & 0 \end{pmatrix}. \tag{33}$$

Table 2. Initial orbital parameters for the moonlets of two triple asteroidal systems

*a) (216) Kleopatra (Descamps et al. 2011)*

| *Orbital parameters* | Alexhelios | Cleoselene |
|---|---|---|
| *Semi-major axis: a* (km) | 678.0 | 454.0 |
| *Eccentricity: e* | 0 | 0 |
| *Inclination: i* (deg) | 51.0 | 49.0 |
| Long. of ascend. node: Ω (deg) | 166.0 | 160.0 |
| Arg. periapsis: *ω* (deg) | 0 | 0 |
| *Mean anomaly: M* (deg) | 0 | 0 |
| *Mass:* (kg) | $4.63 \times 10^{18}$ | $4.67 \times 10^{18}$ |

*b) (153591) 2001 SN263 (Fang et al. 2011)*

| *Orbital parameters* | Beta | Gamma |
|---|---|---|
| *Semi-major axis: a* (km) | 16.633 | 3.804 |
| *Eccentricity: e* | 0.015 | 0.016 |
| *Inclination: i* (deg) | 157.486 | 165.045 |
| Long. of ascend. node: Ω (deg) | 161.144 | 198.689 |
| Arg. periapsis: *ω* (deg) | 131.249 | 292.435 |
| *Mean anomaly: M* (deg) | 248.816 | 212.658 |
| *Density* (g cm$^{-3}$) | 1.0 | 2.3 |

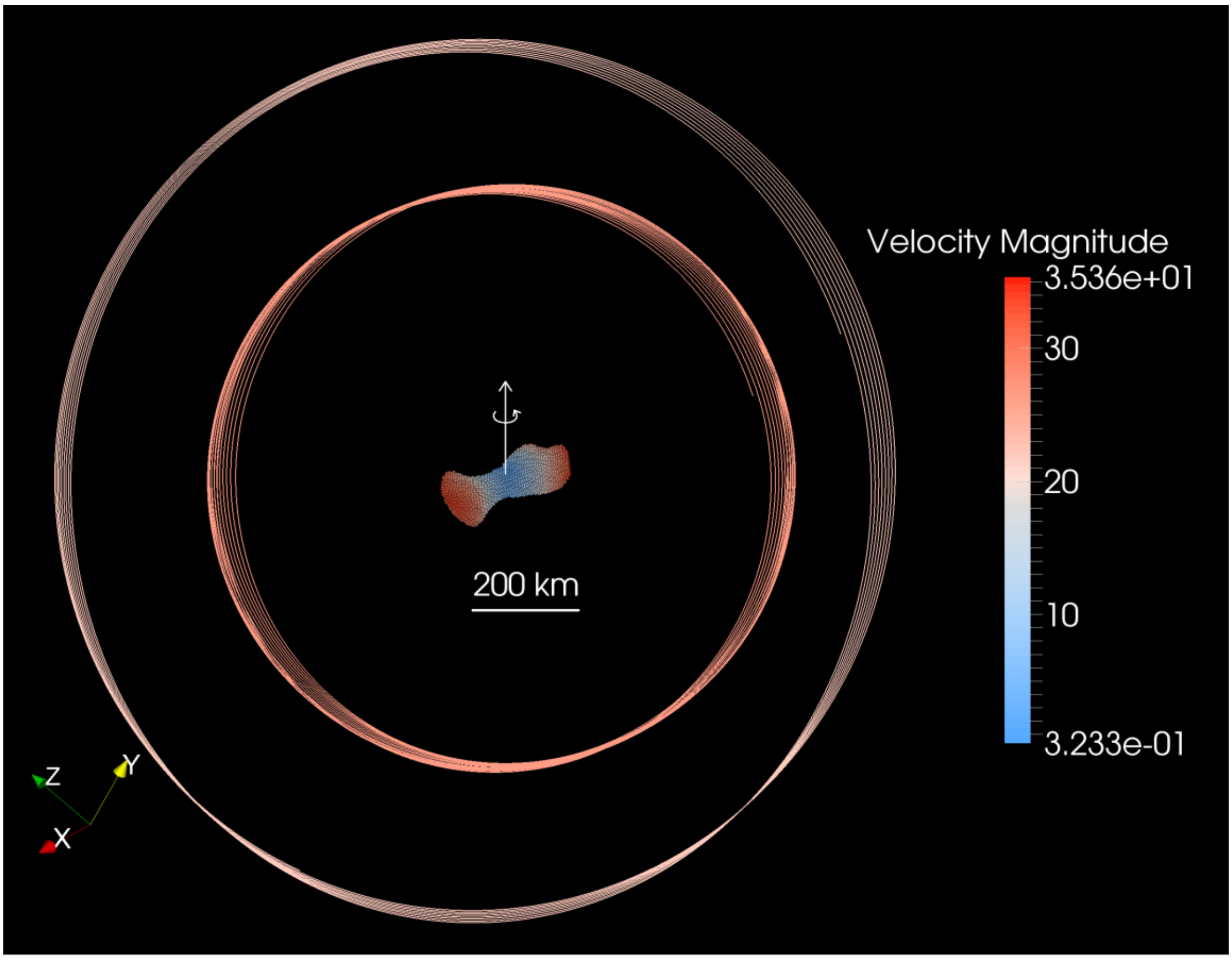

Fig. 1. The dynamical configuration of the two moonlets relative to the primary for the triple asteroidal system (216) Kleopatra, the simulation duration is 28d.

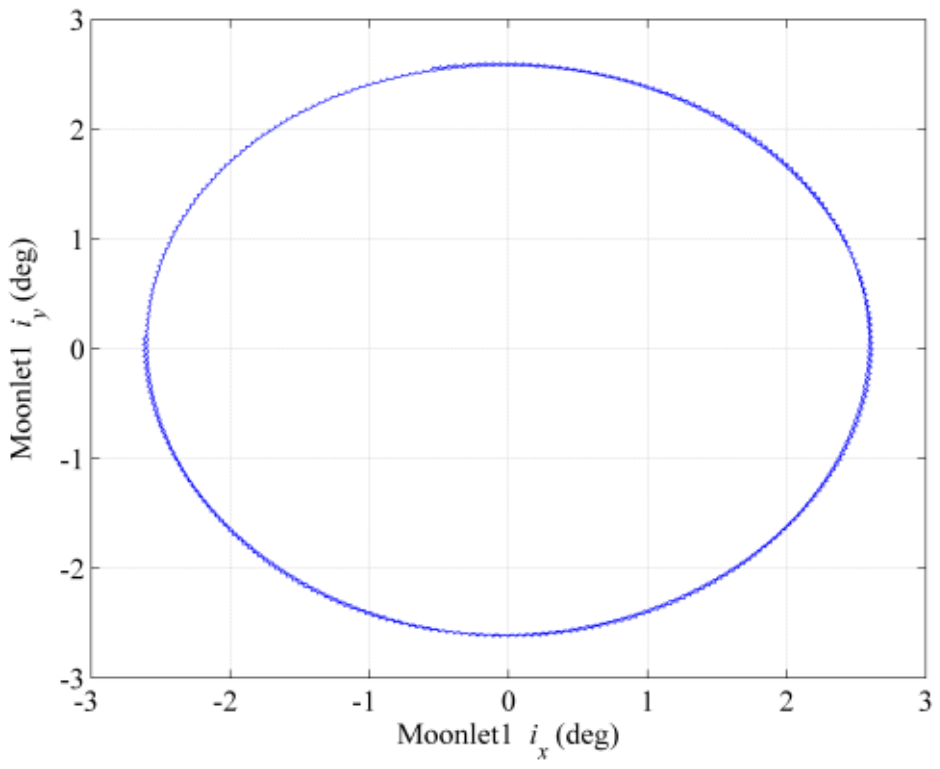

(a)

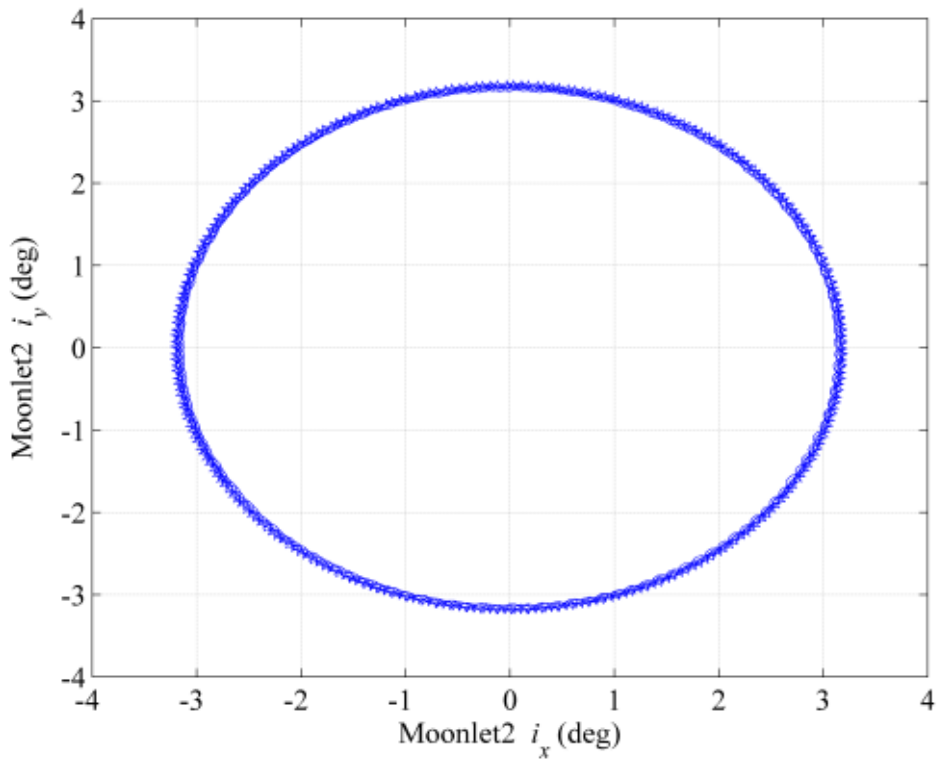

(b)

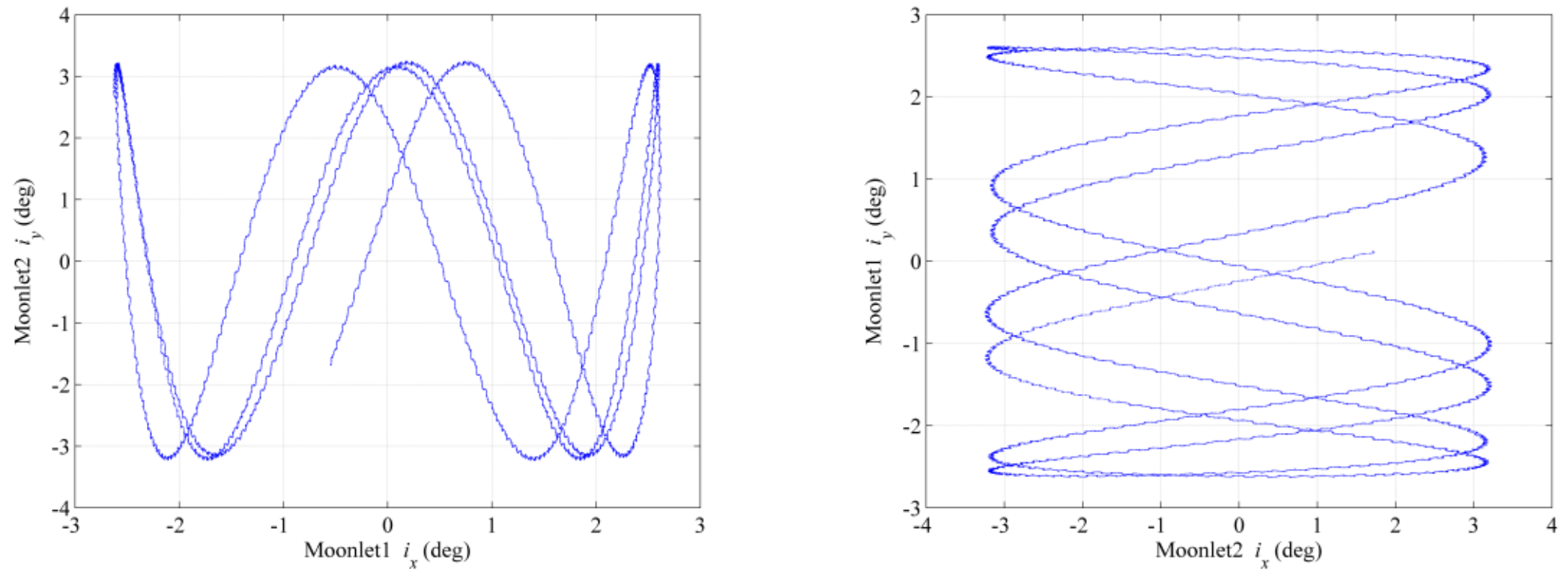

(c) (d)

Fig. 2. The numerical calculation of the components of inclination vectors of two moonlets of the triple asteroidal system (216) Kleopatra, (a) the trajectory of two components of the inclination vector of moonlet 1, (b) the trajectory of two components of the inclination vector of moonlet 2, (c) the trajectory of the x component of the inclination vector of moonlet 1 and y component of the inclination vector of moonlet 2, (d) the trajectory of the x component of the inclination vector of moonlet 2 and y component of the inclination vector of moonlet 1.

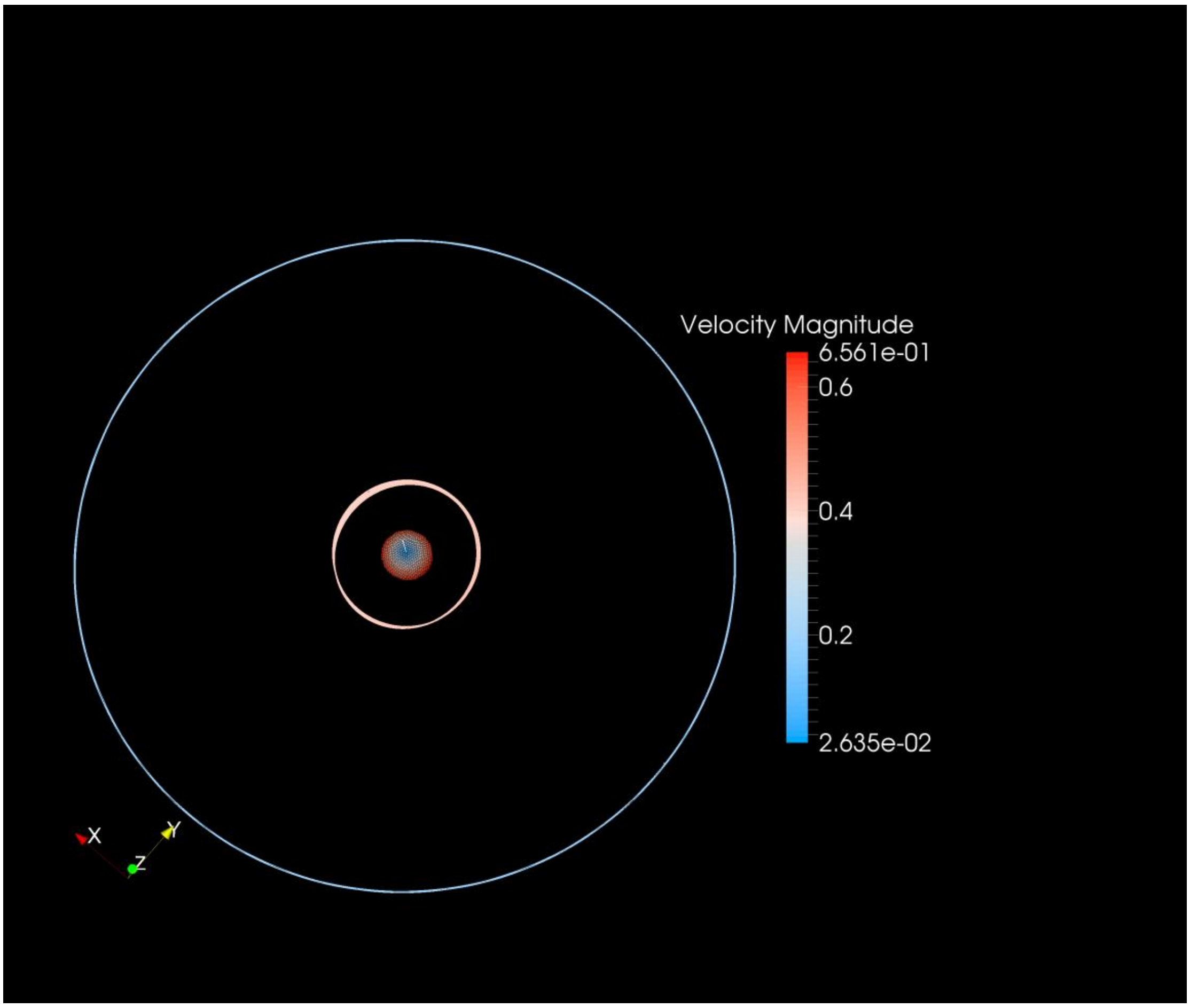

Fig. 3. The dynamical configuration of the two moonlets relative to the primary for the triple

asteroidal system (153591) 2001 SN263 , the simulation duration is 600d.

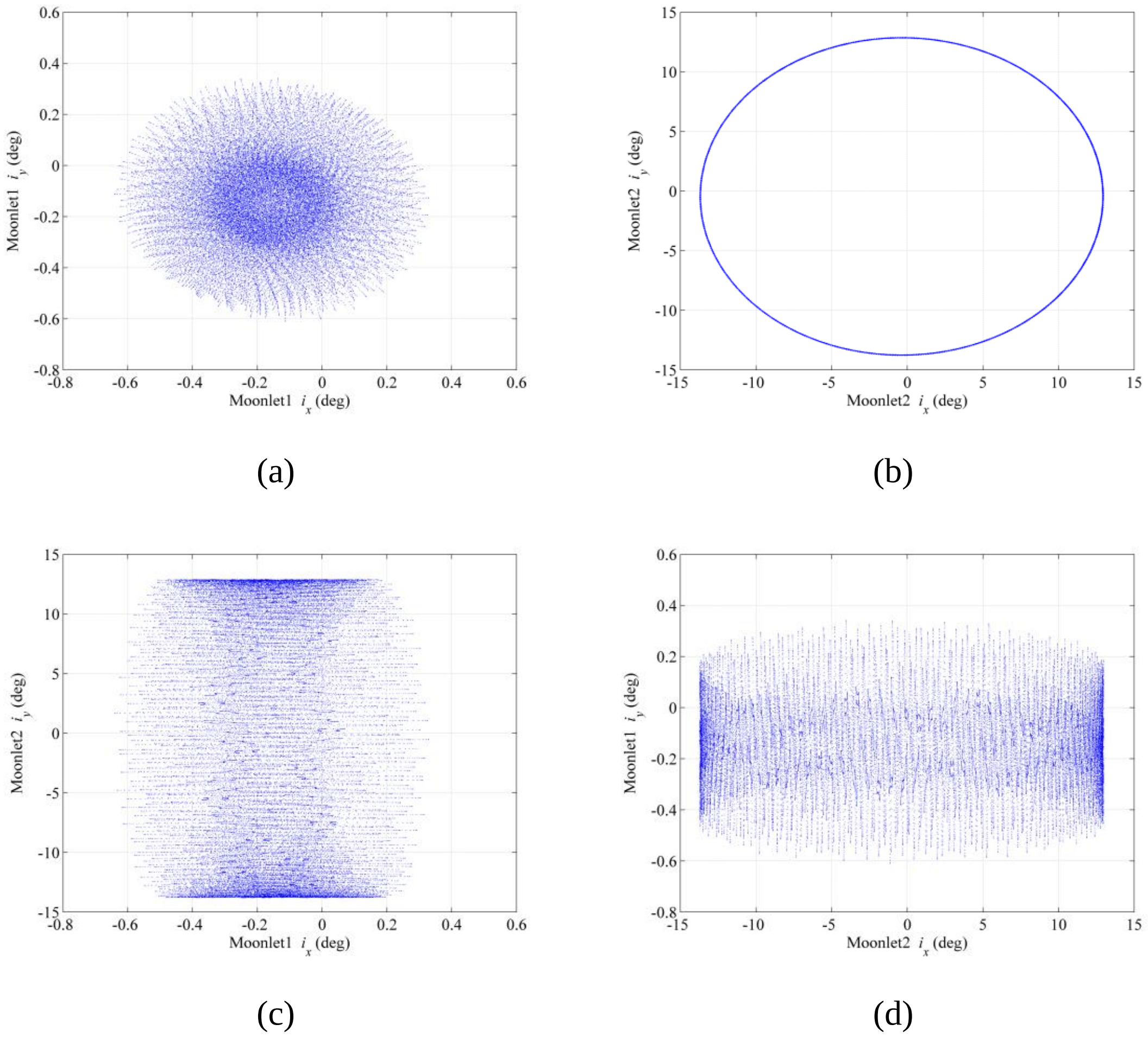

Fig. 4. The numerical calculation of the components of inclination vectors of two moonlets of the triple asteroidal system (153591) 2001 SN263, (a) the trajectory of two components of the inclination vector of moonlet 1, (b) the trajectory of two components of the inclination vector of moonlet 2, (c) the trajectory of the x component of the inclination vector of moonlet 1 and y component of the inclination vector of moonlet 2, (d) the trajectory of the x component of the inclination vector of moonlet 2 and y component of the inclination vector of moonlet 1.

Descamps et al. (2011) presented the orbit parameters of two moonlets of (216) Kleopatra in mean J2000 equator, see Table 2. The frame used here is defined as follows, the origin is the mass center of the primary, the xy plane is the equator of the primary, and z axis is the spin axis of the primary. In our frame, the inclinations of Alexhelios and Cleoselene are 2.6 deg and 3.18 deg, respectively. So the inclinations of these two moonlets are small and the results here can be used to analyze the orbits of these two moonlets. Fig. 1 shows the dynamical configuration of the two moonlets relative to the primary for the triple asteroidal system (216) Kleopatra while Fig. 2 presents the components of inclination vectors of two moonlets. From Fig. 2, one can conclude that the mean motion of $i_x$ and $i_y$ of each moonlet forms an ellipse, and the amplitude of the instantaneous motion of the elliptical trajectory for Cleoselene is bigger than for Alexhelios. Besides, Alexhelios' $i_x$ and Cleoselene's $i_y$ form a quasi-periodic motion, and Alexhelios' $i_y$ and Cleoselene's $i_x$ form a quasi-periodic motion.

For the motion near the surface of asteroids like Kleopatra, the perturbation method with low Legendre coefficients can't model the orbital motion accurately. The reason is that the higher order terms of the Legendre coefficients need many iterations to converge (Elipe and Riaguas 2003). Besides, there exists some orbits where the minimal distance between the mass center of Kleopatra and the orbit is smaller than Kleopatra's mean radius (Jiang et al. 2015c). This means that the perturbation method with low Legendre coefficients can't be used to model the motion near the surface of Kleopatra. However, if the orbit is far from the surface of Kleopatra, the perturbation method with low Legendre coefficients can also be used. The ratio of the semi-major axis of the moonlets and the mean radius of Kleopatra are 6.7 and 10 (Descamps et al. 2011). The numerical method uses the polyhedral model to model the gravity of Kleopatra. The numerical results fit the theoretical results well because the orbits are far from Kleopatra, and the mass ratio of the moonlets and Kleopatra are only

$2.87\times10^{-4}$ and $1.32\times10^{-4}$.

Fang et al. (2011) presented the orbit parameters of two moonlets of (153591) 2001 SN263 in mean J2000 equator, see Table 2. In our frame, the inclinations of Beta and Gamma are 0.33 deg and 13.35 deg, respectively. The inclinations of these two moonlets are also small and the results here can be used to analyze the orbits of these two moonlets. Fig. 3 shows the dynamical configuration of the two moonlets relative to the primary for the triple asteroidal system (153591) 2001 SN263 while Fig. 4 presents the components of inclination vectors of two moonlets. Fig. 4 implies that the mean motion of $i_x$ and $i_y$ of each moonlet forms an ellipse, and the amplitude of the instantaneous motion of the elliptical trajectory for Gamma is much bigger than for Beta. The instantaneous motion of the elliptical trajectory for Beta forms an ellipse while the instantaneous motion of the elliptical trajectory for Gamma forms an elliptical disc. Additionally, Beta's $i_x$ and Gamma's $i_y$ form a quasi-periodic motion, and Beta's $i_y$ as well as Gamma's $i_x$ form a quasi-periodic motion. The numerical calculation validates the above theoretical derivation.

The theoretical results say that the mean motion of x component and the y component of the inclination vector of each moonlet forms an ellipse. From Figs. 2(a), 2(b), 4(a), and 4(b), one can see that the inclination vector of each moonlet forms an ellipse. The relative amplitude of the trajectory in Fig. 4(b) is smaller than that in Fig. 4(a), because the inclinations of Beta and Gamma are 0.33 deg and 13.35 deg in the equator of the primary, respectively. Fig. 4(a) shows the inclination vector of Beta while Fig. 4(b) shows the inclination vector of Gamma. In addition, the theoretical results say that the x component of the inclination vector of moonlet 1 and the y component of the inclination vector of moonlet 2 form a quasi-periodic motion, and the x component of the inclination vector of moonlet 2 and the y component of the inclination vector of moonlet 1 form a quasi-periodic motion. From Figs. 2(c), 2(d), 4(c), and 4(d), one can see that the component of the inclination vector between different moonlets are coupled and form a quasi-periodic motion.

The results shown above agree with previous work based on observational data that concluded periodical variety of the orbital parameters of different triple asteroidal systems. Marchis et al (2010) calculated orbital parameters of the triple asteroidal system (45) Eugenia, and found that the inclinations of these two moonlets of (45) Eugenia are about 9 deg and 18 deg relative to the equator of the primary, and have a periodical variety. Fang et al. (2011) calculated the change rate of the argument of pericenter and the longitude of the ascending node for the two moonlets of (153591) 2001 SN263. Our results also indicate that the longitude of the ascending node have a variety. Fang et al. (2012) also investigated the semi-major axis and eccentricity of Remus and Romulus relative to Sylvia of the triple asteroidal system (87) Syivia, and found both of them have a periodical variety, and the variety period are different. The previous studies only consider the inclinations of the moonlets in the triple asteroidal system. However, the moonlets of (216) Kleopatra and (153591) 2001 SN263 are in the orbit which is near the equator of the primary, the inclination vector is much better to analyze the motion of these moonlets than inclination of these moonlets (Hintz 2008). Fig. 2 and Fig. 4 present the coupling motion of the inclination vector of two moonlets in the triple asteroidal systems.

## 4. Conclusions

The nonlinear dynamical behaviours in the triple asteroidal systems are complicated. The primary has irregular shapes and the moonlets have relative effect. The primary's non-sphericity gravity, the solar gravity, and moonlets' relative gravity are all considered in this paper. It is found that the inclination vector for each moonlet forms a periodic elliptical motion. The Solar gravity and moonlets' relative gravity lead to the periodic motion for $i_x$ of moonlet 1 and $i_y$ of moonlet 2, and the periodic motion for $i_y$ of moonlet 1 and $i_x$ of moonlet 2. The mean motion of $i_x$ and $i_y$ of the inclination vector of each moonlet forms an ellipse. The instantaneous motion of $i_x$ and $i_y$ may be elliptical due to the coupling effect of these forces. The coupling effect of these forces also makes $i_x$ of moonlet 1 and $i_y$ of moonlet 2 form a quasi-periodic motion, and $i_x$ of moonlet 2 and $i_y$ of moonlet 1 form a quasi-periodic motion.

The numerical computation of orbital motion of two triple asteroidal systems (216) Kleopatra and (153591) 2001 SN263 further illustrates the results. The numerical results are compared with the research in existing literature. The moonlets of (216) Kleopatra and (153591) 2001 SN263 motion near the equator of the primary, then the inclinations of these moonlets are small. To analyze the motion of these moonlet, using the inclination vector is better than the inclination. We also compare the numerical results with the theoretical results. It is found that the amplitude of the instantaneous motion of the elliptical trajectory for the moonlet Cleoselene is bigger than for the moonlet Alexhelios in the triple asteroidal systems (216) Kleopatra. The instantaneous motion of the elliptical trajectory for Gamma looks like an elliptical disc in the triple asteroidal systems (153591) 2001 SN263.

## Acknowledgements

This research was supported by the National Natural Science Foundation of China (No. 11372150& No. 11572166), the State Key Laboratory of Astronautic Dynamics Foundation (No. 2016ADL-0202) and the National Science Foundation for Distinguished Young Scholars (11525208).

## Appendix A

In this section, we present the derivation of Eq. (11).

For the first moonlet in the nearly circular orbit near the equatorial plane of the primary, the perturbation force on moonlets due to the Solar gravity and the second moonlet's relative gravity (Tremaine et al. 2008) is

$$\mathbf{F} = rn_c^2 \left[ 3\cos\xi \cdot \left( \frac{\mathbf{r}'}{r'} \right) - \frac{\mathbf{r}}{r} \right], \tag{A1}$$

where all the vectors are expressed in the equatorial inertial frame of the first moonlet, $\mathbf{r}$ represents the first moonlet's position vector, $r$ is the norm of $\mathbf{r}$, $\mathbf{r}'$ represents the Sun's position vector or the second moonlet's position vector, $r'$ is the norm of $\mathbf{r}'$, $\xi$ represents the angle between $\mathbf{r}$ and $\mathbf{r}'$. For the Solar gravity, $n_c = n_s$; for the second moonlet's gravity, $n_c^2 = \sigma_{M2} n_{M2}^2$.

The component of $\mathbf{F}$ in the normal direction of the first moonlet's orbital plane is

$$F_n = \mathbf{F}\cdot\mathbf{n}_0 = 3rn_c^2\cos\xi\left(\frac{\mathbf{r}'}{r'}\cdot\mathbf{n}_0\right), \tag{A2}$$

Where

$$\mathbf{n}_0 = \begin{bmatrix} \sin i\sin\Omega \\ -\sin i\cos\Omega \\ \cos i \end{bmatrix},\quad \frac{\mathbf{r}}{r} = \begin{bmatrix} \cos u_{M1}\cos\Omega_{M1} - \sin u_{M1}\sin\Omega_{M1}\cos i_{M1} \\ \cos u_{M1}\sin\Omega_{M1} + \sin u_{M1}\cos\Omega_{M1}\cos i_{M1} \\ \sin u_{M1}\sin i_{M1} \end{bmatrix}, \tag{A3}$$

for the Solar gravity,

$$\frac{\mathbf{r}'}{r'} = \begin{bmatrix} \cos\beta_s \\ \sin\beta_s\cos i_s \\ \sin\beta_s\sin i_s \end{bmatrix}, \tag{A4}$$

while for the second moonlet's gravity,

$$\frac{\mathbf{r}'}{r'} = \begin{bmatrix} \cos\beta_{M2}\cos\Omega_{M2} - \sin\beta_{M2}\sin\Omega_{M2}\cos i_{M2} \\ \cos\beta_{M2}\sin\Omega_{M2} + \sin\beta_{M2}\cos\Omega_{M2}\cos i_{M2} \\ \sin\beta_{M2}\sin i_{M2} \end{bmatrix}. \tag{A5}$$

Thus the perturbation of the inclination vector due to the Solar gravity is

$$\begin{cases} \dfrac{di_{x-M1}}{dt} = \dfrac{F_n\sin\left(\Omega+\omega+M\right)}{rn_{M1}} \\ = \dfrac{3}{2}n_{M1}\left(\dfrac{n_s}{n_{M1}}\right)^2 \sin\beta_s\cos i_s\left(\cos\beta_s\sin i\sin\Omega - \sin\beta_s\cos i_s\sin i\cos\Omega + \sin\beta_s\sin i_s\right) \\ \dfrac{di_{y-M1}}{dt} = \dfrac{F_n\cos\left(\Omega+\omega+M\right)}{rn_{M1}} \\ = \dfrac{3}{2}n_{M1}\left(\dfrac{n_s}{n_{M1}}\right)^2 \cos\beta_s\left(\cos\beta_s\sin i\sin\Omega - \sin\beta_s\cos i_s\sin i\cos\Omega + \sin\beta_s\sin i_s\right) \end{cases}. \tag{A6}$$

Considering that $\sin i \ll 1$, thus

$$\begin{cases} \dfrac{di_{x-M1}}{dt} = \dfrac{3}{4}n_{M1}\left(\dfrac{n_s}{n_{M1}}\right)^2 \sin^2\beta_s\sin 2i_s \\ \dfrac{di_{y-M1}}{dt} = \dfrac{3}{4}n_{M1}\left(\dfrac{n_s}{n_{M1}}\right)^2 \sin 2\beta_s\sin i_s \end{cases}. \tag{A7}$$

The perturbation of the inclination vector due to the second moonlet's gravity is

$$\begin{cases} \frac{di_{x-M1}}{dt} = \frac{F_n \sin(\Omega + \omega + M)}{rn_{M1}} \\ = \frac{3}{2} n_{M1} \left( \frac{n_s}{n_{M1}} \right)^2 (\cos\beta_{M2} \sin\Omega_{M2} + \cos i_{M2} \sin\beta_{M2} \cos\Omega_{M2}) \left[ \sin i \sin\Omega (\cos\beta_{M2} \cos\Omega_{M2} - \cos i_{M2} \sin\beta_{M2} \sin\Omega_{M2}) \right. \\ \left. - \sin i \cos\Omega (\cos\beta_{M2} \sin\Omega_{M2} + \cos i_{M2} \sin\beta_{M2} \cos\Omega_{M2}) + \sin\beta_{M2} \sin i_{M2} \right] \\ \frac{di_{y-M1}}{dt} = \frac{F_n \cos(\Omega + \omega + M)}{rn_{M1}} \\ = \frac{3}{2} n_{M1} \left( \frac{n_s}{n_{M1}} \right)^2 (\cos\beta_{M2} \cos\Omega_{M2} - \cos i_{M2} \sin\beta_{M2} \sin\Omega_{M2}) \left[ \sin i \sin\Omega (\cos\beta_{M2} \cos\Omega_{M2} - \cos i_{M2} \sin\beta_{M2} \sin\Omega_{M2}) \right. \\ \left. - \sin i \cos\Omega (\cos\beta_{M2} \sin\Omega_{M2} + \cos i_{M2} \sin\beta_{M2} \cos\Omega_{M2}) + \sin\beta_{M2} \sin i_{M2} \right] \end{cases} \tag{A8}$$

Considering that $\sin i \ll 1$, we have

$$\begin{cases} \frac{di_{x-M1}}{dt} = \frac{3}{2} n_{M1} \left( \frac{n_s}{n_{M1}} \right)^2 (\cos\beta_{M2} \sin\Omega_{M2} + \cos i_{M2} \sin\beta_{M2} \cos\Omega_{M2}) \sin\beta_{M2} \sin i_{M2} \\ \frac{di_{y-M1}}{dt} = \frac{3}{2} n_{M1} \left( \frac{n_s}{n_{M1}} \right)^2 (\cos\beta_{M2} \cos\Omega_{M2} - \cos i_{M2} \sin\beta_{M2} \sin\Omega_{M2}) \sin\beta_{M2} \sin i_{M2} \end{cases} \tag{A9}$$

Neglecting the short-term of $\beta_{M2}$, we have

$$\begin{cases} \frac{di_{x-M1}}{dt} = \frac{3}{8} \sigma_{M2} n_{M1} \left( \frac{n_{M2}}{n_{M1}} \right)^2 \sin 2i_{M2} \cos\Omega_{M2} \\ \frac{di_{y-M1}}{dt} = -\frac{3}{8} \sigma_{M2} n_{M1} \left( \frac{n_{M2}}{n_{M1}} \right)^2 \sin 2i_{M2} \sin\Omega_{M2} \end{cases} . \tag{A10}$$